\RequirePackage{fix-cm}
\documentclass[smallextended]{svjour3}       
\smartqed  
\usepackage{graphicx}
\usepackage{subfigure}
\usepackage{mathptmx}      
\usepackage{latexsym}
\usepackage[round]{natbib}
\usepackage{floatrow}

\begin{document}
	
	\title{On The Synergy Between Ariel And Ground-Based High-Resolution Spectroscopy}
	
	

	\author{%
		Gloria Guilluy $^{(1,2)}$ \and Alessandro Sozzetti $^{(1)}$\and Paolo Giacobbe$^{(1)}$  \and Aldo S. Bonomo $^{(1)}$ \and Giuseppina Micela $^{(3)}$
	}
	\authorrunning{Gloria Guilluy \and Alessandro Sozzetti \and Paolo Giacobbe \and Aldo S. Bonomo \and Giuseppina Micela}
	
	\institute{  G. Guilluy  \\
		\email{gloria.guilluy@inaf.it}            \\
		\\
		(1) INAF – Osservatorio Astrofisico di Torino, Via Osservatorio 20, I-10025 Pino Torinese, Italy\\
		(2) Dipartimento di Fisica, Università degli Studi di Torino, via Pietro Giuria 1, I-10125 Torino, Italy \\
		(3) INAF – Osservatorio Astronomico di Palermo, Piazza del Parlamento, 1, 90134, Palermo, Italy\\
	}
	
	\date{Received: date / Accepted: date}
	\maketitle		
	\begin{abstract}
		Since the first discovery of an extra-solar planet around a main-sequence star, in 1995, the number of detected exoplanets has increased enormously. Over the past two decades, observational instruments (both onboard and on ground-based facilities) have revealed an astonishing diversity in planetary physical features (i. e. mass and radius), and orbital parameters (e.g. period, semi-major axis, inclination).
		Exoplanetary atmospheres provide direct clues to understand the origin of these differences through their observable spectral imprints. In the near future, upcoming ground and space-based telescopes will shift the focus of exoplanetary science from an era of “species discovery” to one of “atmospheric characterization”. In this context, the Atmospheric Remote-sensing Infrared Exoplanet Large (Ariel) survey, will play a key role. As it is designed to observe and characterize a large and diverse sample of exoplanets, Ariel will provide constraints on a wide gamut of atmospheric properties allowing us to extract much more information than has been possible so far (e.g. insights into the planetary formation and evolution processes). The low resolution spectra obtained with Ariel will probe layers different from those observed by ground-based high resolution spectroscopy, therefore the synergy between these two techniques offers a unique opportunity to understanding the physics of planetary atmospheres. In this paper, we set the basis for building up a framework to effectively utilise, at near-infrared wavelengths, high-resolution datasets (analyzed via the cross-correlation technique) with spectral retrieval analyses based on Ariel low-resolution spectroscopy. We show preliminary results, using a benchmark object, namely HD~209458~b, addressing the possibility of providing improved constraints on the temperature structure and molecular/atomic abundances.
	\end{abstract}
	
	\section{Introduction}
	\label{intro}
	Past observations have revealed the existence of an amazing diversity in planetary and host star parameters, and beyond all doubt in the properties of exoplanetary atmospheres. 
	An ever-growing number of well-characterized planetary atmospheres -e.g. constrained chemical composition and temperature-pressure (T/P) profile- will permit us to address the origin of these differences and could provide important clues on the formation and evolutionary mechanisms (e.g., \citealt{Madhusudhan_2019} and references therein). In this context, past and current space-borne low-resolution
	spectroscopy (hereafter LRS; at a resolving power of R~$\leq$200), and ground-based high-resolution spectroscopy (hereafter HRS, resolving power R$\geq$20~000) have provided reliable and strong pioneering results. 
	The bulk of atmospheric observations has been accomplished by exploiting the opportunities offered by the class of transiting hot Jupiters. Thanks to their high equilibrium temperatures, and large radii, they were indeed recognized as suitable targets to perform atmospheric detections. Eighteen years of LRS increased our understanding of exoplanetary atmospheres. In the near-infrared (nIR), the Wide Field Camera 3 onboard the Hubble Space Telescope (HST/WFC3) 
	provided transmission spectra for tens of hot Jupiters and for some Sub-Neptunes \citep[e.g.][]{kreidbergGJ, Sing2016, Tsiaras2018, Kreidberg2019, Tsiaras2019}, thus allowing the retrieval of important atmospheric information, such as the molecular composition. In the visible (VIS) band, 
	the Space Telescope Imaging Spectrograph (HST/STIS) and the Advanced Camera for Surveys (HST/ACS), led to the detection of both optical slope \citep[e.g.][]{Lecavelier2008} -which may be due to Rayleigh scattering by molecular hydrogen and/or aerosols- and optical species, such as K and Na, \citep[e.g.,][]{Sing2016,Essen2020}, in several transmission spectra.
	Robust frameworks were then employed to interpret LRS data and derive atmospheric properties, e.g. the Non-linear Optimal Estimator for MultivariatE Spectral analySIS \citep[NEMESIS, ][]{nemesis}, the CaltecH Inverse ModEling and Retrieval Algorithms  \citep[CHIMERA, ][]{Line2013b, Line2014} and the Tau Retrieval for Exoplanets \citep[TauREx, ][]{Waldmann_2015_2, Waldmann_2015,al-refaie_taurex3}. 
	One limitation of the LRS analysis is that, when multiple species overlap, due to the low resolving power, it becomes difficult to determine the contribution of each molecule present in the exo-atmosphere we are dealing with.
	Moreover, the narrow spectral coverage (1.1-1.7~$\mu$m) of the HST/WFC3/G141 grism, mainly allowed for H$_2$O detection, while the abundances of other molecules (like CH$_4$, NH$_3$, HCN, CO, CO$_2$) often remained unconstrained. \\
	More recently, pioneered by \citet{Snellen2010}, a new method has been introduced to characterize exoplanetary atmospheres, namely the HRS. Working with HRS data has two big advantages. First of all, molecular features are resolved into a deep forest of individual lines, a spectrum, which represents the fingerprint of the considered molecule. Thus, at HRS we can reveal the presence of a specific molecule, despite overlap, thanks to the cross-correlation of empirical data with model templates \citep{Birkby2018}. Secondly, during transit, the radial component of the planet orbital motion varies by tens of km~s$^{-1}$ across a few hours, and this allows us to distinguish the planetary spectrum from the telluric and stellar photosphere contamination, which are nearly stationary signals during one observing night. However, sometimes strong telluric residuals can remain in the data biasing the interpretation of the atmospheric composition and properties \citep[e.g.,][]{Brogi_2018_Giano}. On the contrary, LRS observations, as they are mainly gathered with space telescopes, are not affected by telluric lines. \\
	HRS from the ground and LRS from space are complementary, but they are difficult to combine. For instance, as the HRS data are self-calibrated (i.e. by fitting the flux in each spectra channel as a function of time and dividing by the fit), any variation due to the planetary atmosphere is being measured with respect to a local `pseudo-continuum', and the absolute level of the planetary absorption is lost. 
	The lack of a robust continuum  can thus introduce degeneracies, because model templates with different abundances and T/P
	profiles may look similar \citep{Birkby2018}. Consequently, the removal of the continuum makes the typical retrieval framework for the analysis of LRS data unusable, and, therefore, putting strong constraints on both the molecular abundances and the temperature/pressure structure is harder. Furthermore, another trouble in using retrieval algorithms with HRS data is converting the cross-correlation values into a goodness-of-fit estimator. The first attempt at combining LRS and HRS was performed by \citet{Brogi2017}, who employed a new retrieval algorithm able to compute the joint probability distribution of LRS and HSR data. Their results showed that the combination of these two techniques improves constraints on both the vertical thermal structure and the retrieved molecular abundances of a planetary atmosphere. However, this algorithm requires significant computational power, thus its application is limited to the evaluation of a few thousand model HRS spectra sampled from an LRS posterior.  More recently, \citet{Brogi2019} introduced a new approach to solving these restrictions. They built a new robust and unbiased framework to perform Bayesian retrieval analyses on HRS data. This framework is based on the cross-correlation between models to extract the planetary spectral signal. In this way, it permits to combine and explore the synergies between HRS and LRS. However, \citet{Brogi2019} applied this framework to a narrow spectral coverage -i.e. the VLT CRIRES K-band - while, as current spectrographs like GIANO-B, CARMENES, SPIROU, and upcoming such as NIRPS and CRIRES+, have broader wavelength ranges, they would require major computational efforts.\\
	In this work, we present a case study of a representative hot Jupiter, namely HD~2094\-58~b, with the objective of analyzing both HRS data acquired with the spectrograph GIANO-B mounted on the Telescopio Nazionale Galileo, TNG, (Sect.~\ref{GIANO-B}), and LRS data gathered with the HST/WFC3 instrument (Sect.~\ref{sec:1}). More precisely, we carried out the analysis at low resolution by taking as reference the results obtained from our simulation at high resolution. After comparing the results obtained with the two different methods, in Sect.~\ref{ARIEL_sim} we perform a simulation to test the Atmospheric Remote-sensing Infrared Exoplanet Large-survey
	(hereafter, Ariel, \citealt{Tinetti2018}) capability to probe exoplanetary atmospheres. Finally, we conclude (Sect.~\ref{Conclusion}) by highlighting the importance of a future synergy between space-borne low-resolution telescope, like Ariel, and ground-based high-resolution facilities. 
	\section{Observations and analysis of GIANO-B high-resolution spectra} \label{GIANO-B}
	The HRS dataset has been analysed in \citet{Giacobbe2020}. It encompassed 4 transit events gathered with the TNG telescope in the GIARPS observing mode \citep{Claudi2017}, that allows for a simultaneous acquisition of high-resolution spectra in the optical (0.39-0.69~$\mu$m) and in the nIR (0.95-2.45~$\mu$m) with the HARPS-N (resolving power R$\sim$115,000), and the GIANO-B  (resolving power R$\sim$50,000) spectrographs, respectively. 
	We observed the system HD~209458 as part of the Large Program “GAPS2: the origin of planetary systems diversity” [PI: G. Micela]. 
	For the work presented here, we focused on the nIR observations collected with GIANO-B.
	This nIR spectrograph acquires images with the nodding acquisition mode ABAB to enable an optimal subtraction of the background and detector noise. The detailed description of the analysis we carried out on these spectroscopic observations of HD~209458~b is described in \citet{Giacobbe2020}. In this manuscript, for completeness, we summarize the most important steps we performed to extract the planetary signal from the GIANO-B spectra.
	After the extraction and the wavelength calibration of the 50 GIANO-B orders with the GOFIO instrument pipeline \citep{Rainer2018}, we proceeded to eliminate the stellar and telluric contamination. To do it, we took advantage of the fact that the planetary signal is Doppler shifted during one night of observation -the planet is moving around its host star-, whereas the stellar and telluric spectra are stationary (or quasi-stationary) signals in wavelength. In this work, we employed the Principal Component Analysis (PCA) and linear regression techniques to model and remove the telluric and stellar signals. (For more details on this new analysis framework see \citealt{Giacobbe2020}). Successively, we looked for HD~209458~b's atmospheric signatures by cross-correlating our GIANO-B spectra (now free from the telluric and stellar contamination) with model templates. 
	The theoretical spectra have been generated with the GENESIS model \citep{sid}. Our models spanned a wide range of pressures (10$^{2}$-10$^{-8}$~bar) and wavelengths (0.9-2.6~$\mu$m). Collision-induced absorption from H$_2$-H$_2$ and H$_2$-He was included. An isothermal atmosphere and constant Volume Mixing Ratios (VMR) were assumed within the range 1,000$<$T$<$ 1,500~K and 10$^{-5}<$VMR$<10^{-2}$. We utilised the ExoMol database for H$_2$O, NH$_3$, HCN, and C$_2$H$_2$ \citep{polyansky_h2o,44,HCN,acety}, the HITEMP database for CH$_4$ and CO \citep{CH4_hitemp,li_co_2015}, and the Ames database for CO$_2$ \citep{Ames}.
	For each theoretical model, to maximize the planetary signal, we chose to perform the cross-correlation technique on a subset of GIANO-B orders, that is those that did not exhibit strong telluric residuals and contained the strongest spectral lines of the planet spectrum. The cross-correlation functions of these selected orders were then co-added in time as a function of the planetary radial velocity in the Earth's rest frame and maximum radial velocity semi-amplitude (K$_{\mathrm{p}}$).
	For all species, we calculated the detection significance by performing a Welch t-test \citep{welch} , see e.g. \citealt{brogi_tauboo_2012}, on the cross-correlation values by assuming as null hypothesis that out-of-trail (far from the planet radial velocity) and in-trail (around the planet radial velocity) values have the same mean. We assumed a confidence level limit of 3$\sigma$.
	Table~\ref{table0} lists the molecules we detected (H$_2$O, NH$_3$, HCN, C$_2$H$_2$, CH$_4$, and CO) and the corresponding abundances used for the cross-correlation between atmospheric templates and our GIANO-B spectra. We need to underline that these abundances do not match any specific chemical scenarios, but they were used to maximise the detection significance. We did not find any evidence of CO$_2$. Thus, we discarded this molecule from the rest of the analysis carried out in this paper.\\
	To better characterize the atmosphere of HD~209458~b, in the work presented in \citet{Giacobbe2020}, we computed two other sets of
	non-isothermal atmospheric models. For the first set of models, we used input temperature-pressure abundance
	(T-p-VMR) profiles calculated under the assumptions of a cloud-free atmosphere
	in chemical and radiative equilibrium. The second set of
	models accounts for the presence of clouds/aerosol by adding a grey cloud deck with a
	top-deck pressure of 10$^{-5.5}$~bar and a cloud fraction of 0.4 \citep{Barstow2020}. Following the receipt proposed by \citet{Brogi2019}, we thus converted the cross-correlation values into likelihood values, and we used the likelihood-ratio test to compare the different models. Our results statistically favour the presence of aerosols in the atmosphere of HD~209458~b which dampen the amplitude of the molecular lines but do not evidently
	hamper their detection \citep{Gandhi2020,Hood2020}. Furthermore, the atmospheric models we tested in thermochemical equilibrium seem to prefer a carbon-to-oxygen ratio close to or greater than 1, higher than the solar value (0.55).
	\begin{table*}
		\centering
		\caption{Molecular detections (and the correspondent significance) in the atmosphere of HD~209458~b obtained by cross-correlation with isothermal models. From left to right, we report the investigated molecules, the theoretical abundances employed for the isothermal models, and the significance. These abundances do not match any specific chemical scenarios, but they were used to maximise the detection significance.} 
		\begin{tabular}{l c c}
			\hline
			\textbf{Molecule} & \textbf{Abundance} & \textbf{Significance [$\sigma$]}\\
			\hline
			H$_2$O & 1.2$\times$10$^{-4}$ & 9.6\\
			CO     & 1.4$\times$10$^{-3}$ & 5.5\\
			HCN    & 8.6$\times$10$^{-5}$ & 9.9\\
			CH$_4$ & 4.7$\times$10$^{-3}$ & 5.6\\
			NH$_3$ & 1.3$\times$10$^{-4}$ & 5.3\\
			C$_2$H$_2$ & 8.3$\times$10$^{-5}$ & 6.1\\
			\hline
		\end{tabular}
		\label{table0}
	\end{table*}
	
	\section{Observations and analysis of HST/WFC3 low-resolution data} \label{sec:1} 
\begin{figure}
		\centering
		\includegraphics[height=8cm,width=\linewidth]{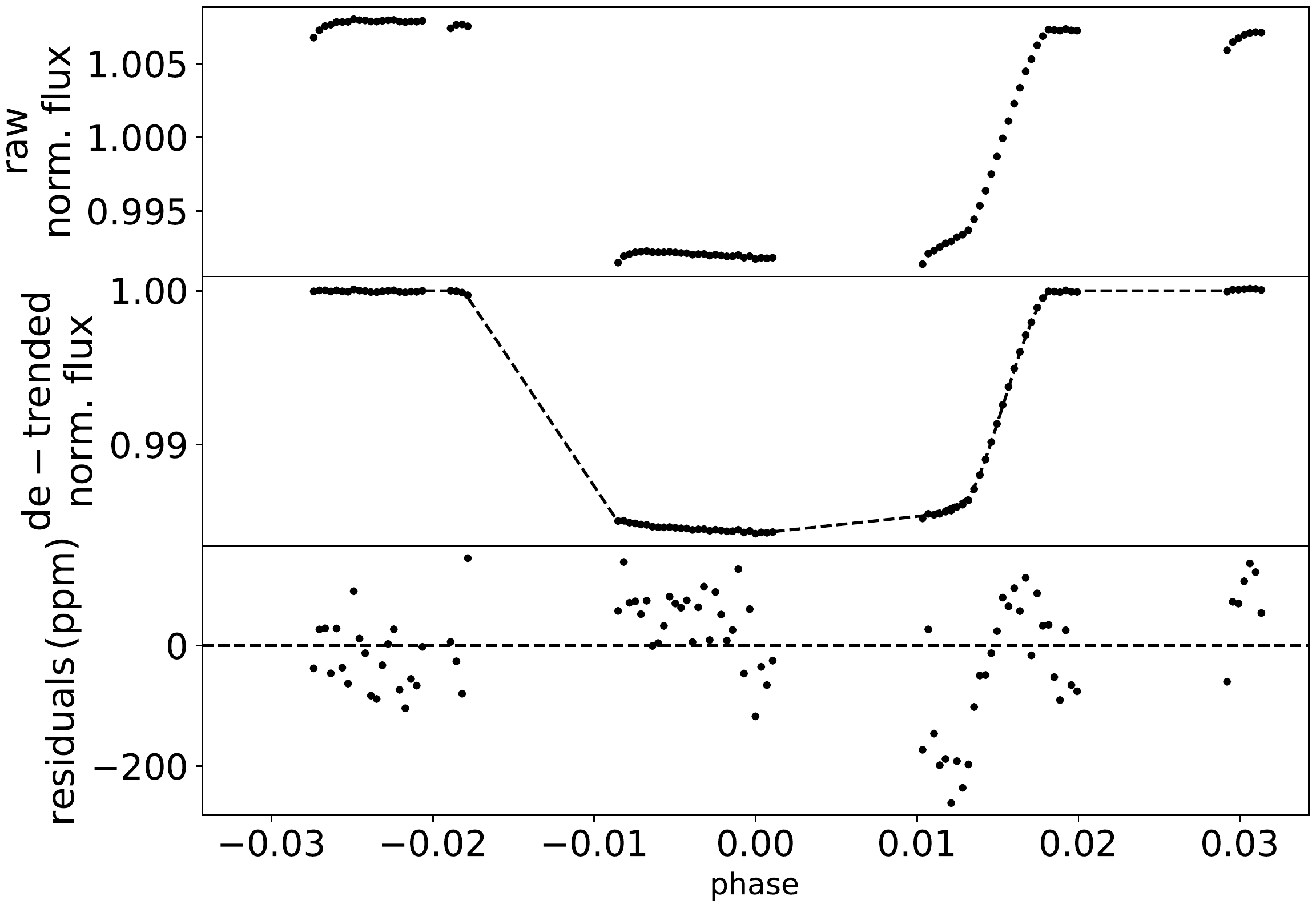}
		\caption{White light curves obtained from \texttt{Iraclis}. Top panel:  normalized raw light-curve.  Second panel:light-curve divided by the best-fit model for the systematics.  Third panel:  Fitting residuals. 
			As in \citet{Tsiaras2016}, we note additional residuals in the white light curve fit. The model fails to fit the egress. In this previous work, they attributed this behavior to either non-optimal values
			used for i and a/R$_\star$  or remaining
			systematics.}
		\label{WLC}       
	\end{figure}

\begin{figure}
		\centering
		\includegraphics[height=8cm,width=\linewidth]{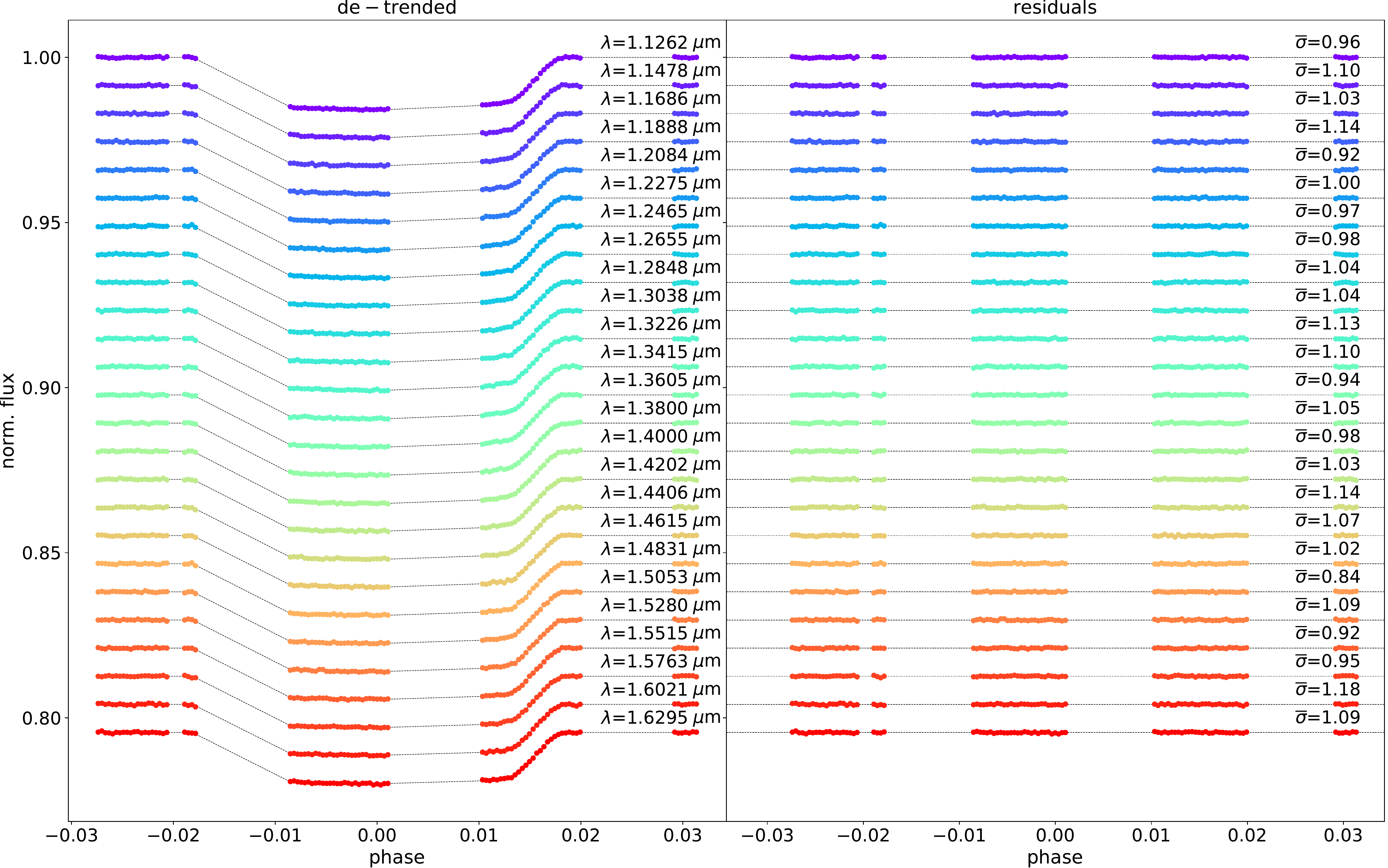}
		\caption{Spectral light curves obtained from \texttt{Iraclis} plotted with an offset for clarity. Left: the detrended light curves with best-fit model plotted. Right panels: residuals, $\bar{\sigma}$ indicates the ratio between the standard deviation of the residuals and the photon noise.}
		\label{fig2}       
	\end{figure}

	The LRS observations employed in this work have been acquired with the HST/WFC3\-/G141 grism as part of the 12181 program, `The Atmospheric Structure of Giant Hot Exoplanets' [PI: D. Deming]. We downloaded publicly available observations of HD~209\-458~b from the Mikulski Archive for Space Telescopes (MAST).
	We used the publicy available Python package \texttt{Iraclis} \citep{Tsiaras2018} to analyze the raw  HST spatially scanned spectroscopic images and to extract the planetary spectrum.  We then carried out the modeling of the extracted spectrum using the publicly available spectral retrieval algorithm TauREx3 \citep{Waldmann_2015_2, Waldmann_2015,al-refaie_taurex3}.
	\\
	The pipeline \verb+Iraclis+ \citep{Tsiaras2018} is composed of different modules: (i) data reduction and calibration; (ii) light curves extraction; (iii)
	limb-darkening coefficients calculation; (iv) white light curves fitting; (v) spectral light curves fitting. Step (i) consists in bias-level and zero-read corrections; non-linearity correction; dark current subtraction; gain variations correction; sky background subtraction; calibration; flat-field correction; bad pixels and cosmic ray correction \citep{Tsiaras2016b, Tsiaras2016, Tsiaras2018}. 
	After these initial operations, the stellar flux was extracted from the raw images to create the wavelength-dependent light curves. Two types of light curves were extracted, a broad wavelength band white light curve covering the entire G141 grism wavelength range (1.088 - 1.68 $\mu$m, see Fig.~\ref{WLC}) and spectral light curves with a resolving power of 70 at 1.4~$\mu$m (see Fig.~\ref{fig2}).
	When extracting the spectral light curves, \verb+Iraclis+ accounts for two “ramps", i.e. time-dependent systematics introduced by the WFC3/IR detector. The first ramp has a linear behavior and affects each HST visit, while the second one alters each orbit and has an exponential behavior. To correct for these effects, the white light curve is fitted by \verb+Iraclis+ with a model for the systematics simultaneously with the transit model. Since, in this data set, the long-term ramp can be approximated by a linear function only after the third orbit, following \citet{Tsiaras2016}, we decided to discard the first two orbits. We fitted for T$_0$ (the mid-transit time) and R$_{\rm {P}}$/R$_\star$ (the ratio of planet to stellar radius) as free parameters, we fixed the orbital period (P), the inclination (i) and a/R$_\star$ to the values of Table~\ref{table1}, and assumed a circular orbit given that the eccentricity is consistent with zero (e.g., \citealt{Bonomo2017}). We modeled the stellar limb darkening effect by employing the non-linear formula with four terms by \citet{2000A&A...363.1081C}. The coefficients are calculated by fitting the stellar profile from an ATLAS model \citep{1970SAOSR.309.....K,2011MNRAS.413.1515H} and by using the stellar parameters presented in Table~\ref{table1}. We then fitted the spectral light curves by using the dividing method proposed by \cite{KreidbergB2014b}. This technique considers the white light curve as a comparison source, indeed each spectral light curve is fitted with a model that includes the white light curve and its best-fit model. This has as consequence that the residuals from fitting one of the spectral light curves do not show trends similar to those in the white light curve. In this fitting procedure, the only free parameter is R$_{\rm {P}}$/R$_\star$, while the other parameters are the same employed for the white light curve fitting.
	The transmission spectrum is then constructed from the spectral light curves by determining the planet-star radii ratio as a function of wavelength (see Table~\ref{table3}).
	\begin{table}[]
		\caption{Planetary, transit, and stellar parameters adopted in this work.}
		\centering
		\begin{tabular}{c c c}
			\hline
			\textbf{Stellar Parameters} & &  \\
			Teff (K) & 6065$\pm$50 & \cite{Torres2008} \\
			$[$Fe/H$]$ (dex) & 0.00$\pm$0.05 & \cite{Torres2008} \\
			R$_{\star}$ (Re) & 1.155$\pm$0.016 & \cite{Torres2008} \\
			log(g$_{\star}$) (cgs) & 4.361$_{-0.008}^{+0.007}$ & \cite{Torres2008} \\
			\hline
			\textbf{Planetary Parameters} & & \\
			T$_\mathrm{eq}$ (K) & 1484$\pm$18 & \cite{Evans2015} \\
			M$_{\rm{P}}$ (M$_{\rm{Jup}}$) & 0.685$\pm$0.015 & \cite{Torres2008} \\
			R$_{\rm{P}}$ (R$_{\rm{Jup}}$) & 1.359$\pm$0.019 & \cite{Torres2008} \\
			\hline
			\textbf{Transit Parameters} & &  \\
			T$_0$ (HJD) & 2452826.628521$\pm$0.000087 & \cite{Knutson2007} \\
			Period (days) & 3.52474859$\pm$0.00000038 & \cite{Knutson2007} \\
			R$_{\rm{P}}$/R$_{\star}$ & 0.12086$\pm$0.00010 & \cite{Torres2008} \\
			a/R$_{\star}$ & 8.76$\pm$0.04 & \cite{Torres2008} \\
			i (deg) & 86.71$\pm$0.05 & \cite{Torres2008} \\
			
			\hline
		\end{tabular}
		\label{table1}
	\end{table}

	\begin{table*}
		\label{table3}
		\begin{tabular}{c c c} 
			\hline
			\textbf{$\lambda$[$\mu$m]} & \textbf{Bandwidth[$\mu$m]} & \textbf{(R$_{\rm P}$/R$_\star)^2$} \\
			\hline
			1.12625 & 0.0219 & 0.01458 $\pm$ 0.00003  \\
			1.14775 & 0.0211 & 0.01455 $\pm$ 0.00004  \\
			1.16860 & 0.0206 & 0.01457 $\pm$ 0.00003  \\
			1.18880 & 0.0198 & 0.01454 $\pm$ 0.00004  \\
			1.20835 & 0.0193 & 0.01452 $\pm$ 0.00003  \\
			1.22750 & 0.0190 & 0.01459 $\pm$ 0.00003  \\
			1.24645 & 0.0189 & 0.01450 $\pm$ 0.00003  \\
			1.26550 & 0.0192 & 0.01449 $\pm$ 0.00003  \\
			1.28475 & 0.0193 & 0.01455 $\pm$ 0.00004  \\
			1.30380 & 0.0188 & 0.01453 $\pm$ 0.00003  \\
			1.32260 & 0.0188 & 0.01449 $\pm$ 0.00004  \\
			1.34145 & 0.0189 & 0.01460 $\pm$ 0.00004  \\
			1.36050 & 0.0192 & 0.01467 $\pm$ 0.00003  \\
			1.38005 & 0.0199 & 0.01468 $\pm$ 0.00003  \\
			1.40000 & 0.0200 & 0.01473 $\pm$ 0.00003  \\
			1.42015 & 0.0203 & 0.01461 $\pm$ 0.00003  \\
			1.44060 & 0.0206 & 0.01462 $\pm$ 0.00004  \\
			1.46150 & 0.0212 & 0.01454 $\pm$ 0.00004  \\
			1.48310 & 0.0220 & 0.01462 $\pm$ 0.00003  \\
			1.50530 & 0.0224 & 0.01459 $\pm$ 0.00003  \\
			1.52800 & 0.0230 & 0.01460 $\pm$ 0.00004  \\
			1.55155 & 0.0241 & 0.01454 $\pm$ 0.00003  \\
			1.57625 & 0.0253 & 0.01456 $\pm$ 0.00003  \\
			1.60210 & 0.0264 & 0.01447 $\pm$ 0.00004  \\
			1.62945 & 0.0283 & 0.01456 $\pm$ 0.00004  \\
			\hline
		\end{tabular}
		\caption{Transmission spectrum of HD~209458~b extracted with the \texttt{Iraclis} pipeline.}
	\end{table*}
	
	Successively, we used the Bayesian atmospheric retrieval framework TauREx3 \citep{Waldmann_2015_2, Waldmann_2015,al-refaie_taurex3} to fit the WFC3 spectrum. Thanks to the implementation of the nested sampling code Multinest \citep{Feroz2009}, TauREx allows us to explore the parameter space and find the best fit for the transmission spectrum extracted with \texttt{Iraclis}.  In our retrieval analysis, we used 1000 live points and an evidence tolerance of 0.5.
	We used the plane-parallel approximation to model the atmospheres, with pressures ranging from 10$^{-2}$ to 10$^6$ Pa sampled uniformly in log-space by 100 atmospheric layers.
	The atmosphere is simulated by assuming an isothermal T/P profile (T=T$_\mathrm{eq}$=1484~K, see Table~\ref{table1}) and constant molecular abundances as a function of altitude.  These assumptions are reasonable because, due to the limited wavelength range of the HST/WFC3/G141 grism, we are probing a narrow range of the planetary atmosphere \citep{Tsiaras2018}.
	Considering the results obtained at HRS with GIANO-B (see Sect.~\ref{GIANO-B}), we decided to consider in our fit the following active-gases: H$_2$O \citep[][]{polyansky_h2o}, CH$_4$ \citep[][]{CH4}, CO \citep[][]{li_co_2015},  NH$_3$ \citep[][]{44}, C$_2$H$_2$ \citep[][]{acety}, HCN \citep[][]{HCN}. Each molecular abundance was allowed to vary between 10$^{-12}$ and 10$^{-2}$ in volume mixing ratios as a log uniform prior. We assumed the bulk composition of the atmosphere, similar to that of Jupiter, i.e. made up of a mixture of 85$\%$ hydrogen and 15$\%$ helium with a ratio H$_2$/He=0.17647, 85\% H$_2$ and 15\% He - the Jupiter Chemical composition. 
	We used absorption cross-sections at 15000. 
	Furthermore, we assumed uniform priors for the 10 bar radius (R = 1.3$-$1.4~R$_{\rm{Jup}}$), and the temperature (T = 1000$-$1800~K).  Rayleigh scattering and collision-induced absorption of
	H$_2$–H$_2$ and H$_2$–He \citep{abel_h2-h2,fletcher_h2-h2,abel_h2-he} were also included. We tested two different scenarios by assuming a cloudy atmosphere (model [1] in Table~\ref{Table4_}) and a cloud-free atmosphere (model [2] in Table~\ref{Table4_}). Clouds were fitted assuming a grey opacity model and cloud top pressures —i.e., the pressure at which the cloud starts to be
	opaque- bounds were set between 10$^{-2}$ and 10$^6$~Pa.
	All the priors we assumed are listed in Table~\ref{Table4_}.
	In our analysis, the fitted parameters were the molecular abundances, the temperature, the mean molecular weight, the radius at 10 bar, and, in the cloudy scenario, the cloud top pressure. 
	In agreement with \citet{Tsiaras2018}, we quantified the significance of our detections with the Atmospheric Detection Index (ADI) a  positively defined Bayes Factor between the nominal atmospheric model and a flat-line model (i.e. a model representing a fully cloudy atmosphere, which contains no active trace gases, Rayleigh scattering or collision-induced absorption). This value was then translated into a statistical significance \citep{Kass1995} by using Table~2 of \citet{Benneke2013}.\\
	
\begin{figure}
		\includegraphics[width=\linewidth]{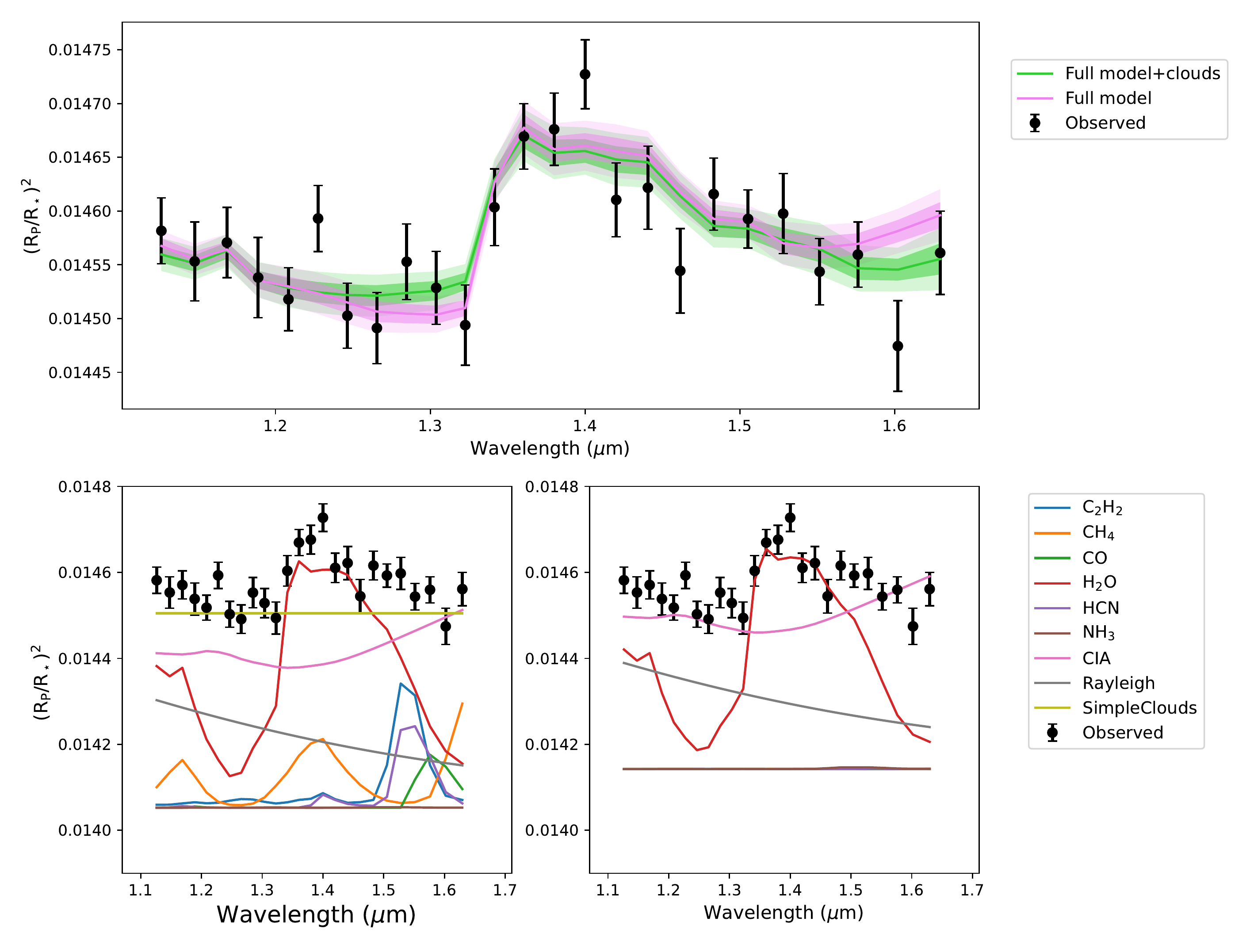}
		\caption{ Upper Panel: Best-fit models for the two different scenarios tested here: a cloudy atmosphere (green), and a cloud-free atmosphere (violet). Bottom Panel: contribution plot for the cloudy-case (left panel) and for the cloud-free scenario (right panel).}
		\label{fig_new}
	\end{figure}

\begin{figure*}
		\centering
		\includegraphics[width=\linewidth]{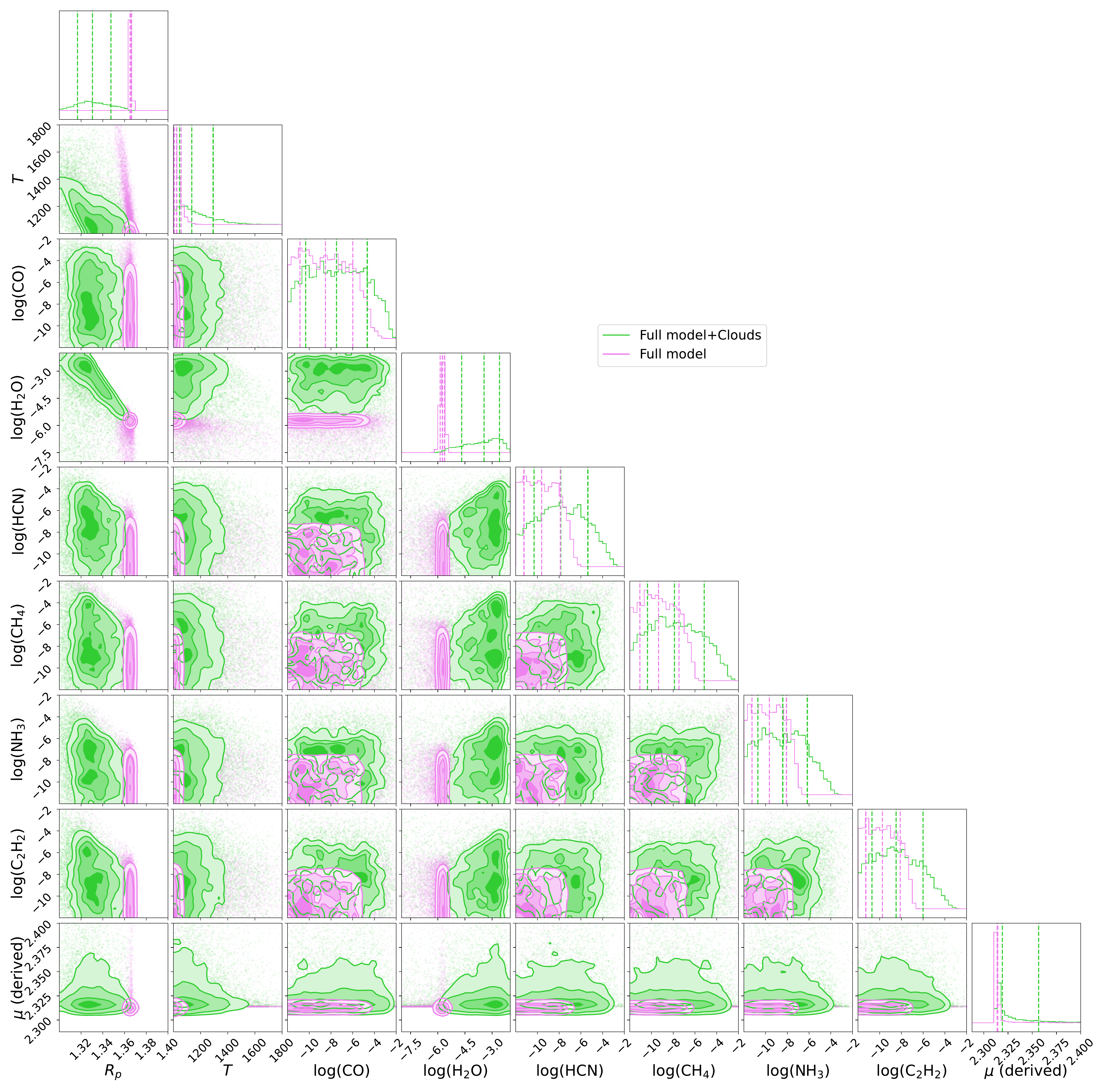}
		\caption{Atmospheric retrieval posterior distributions of real HST/WFC3 observation analysed in this work. The cloudy and the cloud-free scenarios are represented in green and in pink, respectively. For consistency, we do not plot the clouds pressure posterior distribution.}
		\label{fig_4}
	\end{figure*}

	Our TauREx3 retrieval results are listed in Table~\ref{Table4_}. Fig.~\ref{fig_new} shows the best-fit models and the contribution plots for the two scenarios tested. The posterior distributions are shown in Fig.~\ref{fig_4}.
	In both cases, we retrieved a significant atmosphere around HD~209458~b with an ADI of 22.1-17.4 for the cloudy and clear model, respectively.
	Accordingly to the Bayesian evidences the cloudy scenario seems to be strongly, but not decisively, favorite. 
	For both scenarios the retrieved temperature is lower than the predicted equilibrium temperature. This could be explained by the fact that we are probing the atmosphere in the terminator area, and we modeled the atmosphere in 1D using an isothermal profile  \citep[e.g.][]{Caldas2019, MacDonald2020}. This is in agreement with what was highlighted by \citet{Skaf2020}, who found evidence of a global trend between the equilibrium and the retrieved temperatures, with the latter almost always showing lower values. 
	The retrieved radii are compatible with the theoretical value (1.359$\pm$0.019~R$_\mathrm{J}$) within $\sim$1$\sigma$.
	In the cloudy-model, we note a correlation between the H$_2$O abundance, the radius, and the cloud pressure. For less H$_2$O, the model requires deeper clouds and a higher base planet radius.
	Water is required to fit the absorption feature at $\sim$1.4~$\mu$m.
	The posterior distribution of water abundance appears to be much constrained in the clear-scenario case, it reaches its maximum at (log$_{10}$[H$_2$O]= -5.74 $\pm$0.12). 
	The other abundance logarithmic distributions do not seem well constrained, and we can only put upper limits of $10^{-6}$ on the abundance of HCN, C$_2$H$_2$, CH$_4$, NH$_3$. 
	The broad posterior distributions in both the tested scenarios could be due to the combined effect of (i) having a lower spectral coverage compared to that of GIANO-B -at the HST/WFC3 wavelengths, water vapor absorbs much more than the other molecules- and (ii) the possible presence of clouds. 
	Indeed, if in the planetary atmosphere a deep cloud deck is present, as our analysis suggests, it can mask the molecular bands of HCN, NH$_3$, CH$_4$, C$_2$H$_2$, and CO below a pressure level that coincides with the clouds top (log$_{10}$[P$_{\rm clouds}$]=2.07$^{+ 1.28 } _{- 0.82 }$~Pa)\footnote{The retrieved cloud-top pressure is greater than the value reported in the literature (i.e. 10$^{-5.5}$~bar from \citealt{Barstow2020}) and used in Sect~\ref{GIANO-B} to compute the high-resolution theoretical transmission models. This discrepancy could be due to differences in the data reduction pipeline, assumptions made during the retrieval of the real data (e.g., the reference pressure value) or to the spectral range investigated (i.e. the nIR band in this paper and from the optical to the nIR in \citealt{Barstow2020}). Our findings are however in agreement with \citet{Tsiaras2016}, and \citet{Tsiaras2018}.}. This can also be seen in the contribution plot in the bottom-left panel of Fig.~\ref{fig_new}; the yellow line represents the top cloud pressure retrieved by TauREx for the best fit solution.  The signal is theoretically blocked  by  this  layer  and  nothing  can  be  observed at higher pressures.  Molecules found below this line are unconstrained. On the contrary, HRS is able to detect molecular absorbers even if there is a high altitude clouds deck, given that it is most sensitive to the spectral lines cores that are formed above the clouds at lower pressure (see \citealt{Gandhi2020,Hood2020}).
	
	\section{Ariel Spectra Simulation}
	\label{ARIEL_sim}
	\begin{table*}
		\centering
		\resizebox{\columnwidth}{!}{
			\centering
			\begin{tabular}{l c c c c c c c c c c c c} 
				\hline\hline
				\textbf{Model} & & \textbf{H$_2$O} & \textbf{CO} & \textbf{HCN} & \textbf{CH$_4$} & \textbf{NH$_3$} & \textbf{C$_2$H$_2$}  & \textbf{P$_{\rm{clouds}}$} &\textbf{ R$_{\rm{P}}$} & 	\textbf{T$_{\rm {P}}$ } & \textbf{ADI} & \textbf{Log Evidence}\\
				\hline
				& & VMR & VMR & VMR & VMR & VMR & VMR & Pa & R$_{\rm J}$ &  K & \\
				& Scale & log$_{10}$ & log$_{10}$ & log$_{10}$ & log$_{10}$ & log$_{10}$ & log$_{10}$ & log$_{10}$ & linear & linear & \\
				& Priors & [-12;-2] &  [-12;-2] & [-12;-2] &   [-12;-2]  & [-12;-2]  & [-12;-2] & [-2;+6] & [1.3;1.4] & [1000-1800] &   & \\
				\hline
				$[$1$]$  & &  -3.45 $^{+ 0.86 } _{- 1.23 }$ & -7.49 $^{+ 2.83 } _{- 2.84 }$ & -7.84 $^{+ 2.48 } _{- 2.46 }$ & -7.89 $^{+ 2.74 } _{- 2.48 }$ & -8.42 $^{+ 2.26 } _{- 2.28 }$ & -8.49 $^{+ 2.48 } _{- 2.22 }$ & 2.07 $^{+ 1.28 } _{- 0.82 }$ & 1.331 $^{+ 0.017 } _{- 0.014 }$  & 1136 $^{+ 157 } _{- 89 }$ &  22.1 & 211.81\\
				$[$2$]$ & & -5.74 $^{+ 0.12 } _{- 0.12 }$ & -8.51 $^{+ 2.52 } _{- 2.34 }$&  -9.62 $^{+ 1.72 } _{- 1.62 }$ & -9.36 $^{+ 1.89 } _{- 1.71 }$ &-9.65 $^{+ 1.57 } _{- 1.60 }$ & -9.74 $^{+ 1.64 } _{- 1.53 }$ & / & 1.3659 $^{+ 0.0006 } _{- 0.0007 }$ & 1024 $^{+ 33 } _{- 18 }$ &  17.4 & 207.12\\
				\hline
			\end{tabular}
		}
		\label{Table4_}
		\caption{List of the retrieved parameters, their uniform
			prior bounds, the scaling used and the retrieved value for both the two scenarios tested in this paper (model [1] and model [2]). In the last two columns the ADI index and the log of the Bayesian Evidence are also reported}.
	\end{table*}
	Upcoming observatories in space and on the ground, thanks to their broader spectral coverage and higher signal-to-noise ratio, will enable the analysis of a large number of planets. In this contest, during its 4-yr mission, the ESA/Ariel mission, scheduled to launch in 2028, will allow for the atmospheric characterisation of a statistically significant population of exoplanets ($\sim$1000) -ranging from
	Jupiters and Neptunes down to super-Earths. It will thus permit to constrain the abundances of atmospheric constituents which are strictly linked to the planetary formation/evolution environment.  
	
	In this section, we describe a simulation we performed to explore the potential of using Ariel with the approach described above.
	As in the rest of the paper our benchmark object is HD~209458~b. 
	Firstly, we simulated an high-resolution transmission spectrum of HD~209458~b by using the TauREx3 algorithm in forward mode. 
	We simulated the atmosphere of HD~209458~b composed of H$_2$ and He with a ratio H$_2$/He=0.17647 . We modeled the atmosphere with pressures ranging from 10$^{-2}$ to 10$^6$~Pa, uniformly sampled in log-space with 100 atmospheric layers. We assumed an isothermal T/P profile, with T=1080~K, i.e. the maximum of the temperature posterior distribution obtained from the retrieval of HST/WFC3 observations\footnote{We used the maximum of the temperature posterior distribution and not the median value because the distribution seems to converge towards the lower edge of our priors.} (see Fig.~\ref{fig_new}), and constant chemistry profile.  The planetary mass and stellar parameters (i.e. radius and effective temperature) were fixed to the values reported in Table~\ref{table1}, whilst the 10~bar radius was assumed to be equal to that found in the retrieval analysis of real HST/WFC3 observation, i.e. R$_\mathrm{P}$=1.331~R$_\mathrm{jup}$ (see table~\ref{Table4_}). We included Rayleigh scattering and collision-induced absorption (CIA) of  H$_2$–H$_2$ and H$_2$–He.  We assumed a grey opaque cloud deck at P$_{\rm clouds}\sim$ 118~Pa (value obtained from the retrieval analysis of real HST observations -see table~\ref{Table4_}). The trace gases we considered were the same included in the HST retrieval analysis (i.e. H$_2$O, CO, CH$_4$, NH$_3$, C$_2$H$_2$, and HCN). We performed two different simulations: \\
	(a) A first experiment was run requiring molecular abundances being dictated by one of the models in thermochemical equilibrium maximised by the likelihood framework at HRS. In particular, we choose the model with C/O$\sim$0.9 and log$_{10}$[M/H]$\sim$1, with the following abundances:
	\begin{itemize}
		\item[--]   VMR(H$_2$O)=9.2e-5
		\item[--]	VMR(CO)=7.4e-4
		\item[--]	VMR(HCN)=2.1e-8
		\item[--]	VMR(CH$_4$)=6.9e-7
		\item[--]	VMR(NH$_3$)=6.9e-8
		\item[--]	VMR(C$_2$H$_2$)=1.2e-10
	\end{itemize}
	which are the averaged abundance values in the pressures range probed by HST/WFC3, i.e. 0.01-1~bar (see Extended Data Fig.~3 of \citealt{Giacobbe2020}). \\
	(b) We then tried a second experiment -a more unrealistic example- where we imposed the tracers abundances to be equal to those values that maximise the cross-correlation analysis at HRS (see Table~\ref{table0}). We must emphasize that this simulation is only an exercise, it does not claim to reproduce a real observed HST spectrum. We are in fact using the abundances reported in Table~\ref{table0} which, as previously pointed out, do not correspond to any specific chemico-physical scenario of the atmosphere. The aim of this experiment is to assess Ariel's improvement in abundance retrievals with respect to HST/WFC3 in case of higher abundances.\\
	
\begin{figure}
		\includegraphics[width=\linewidth]{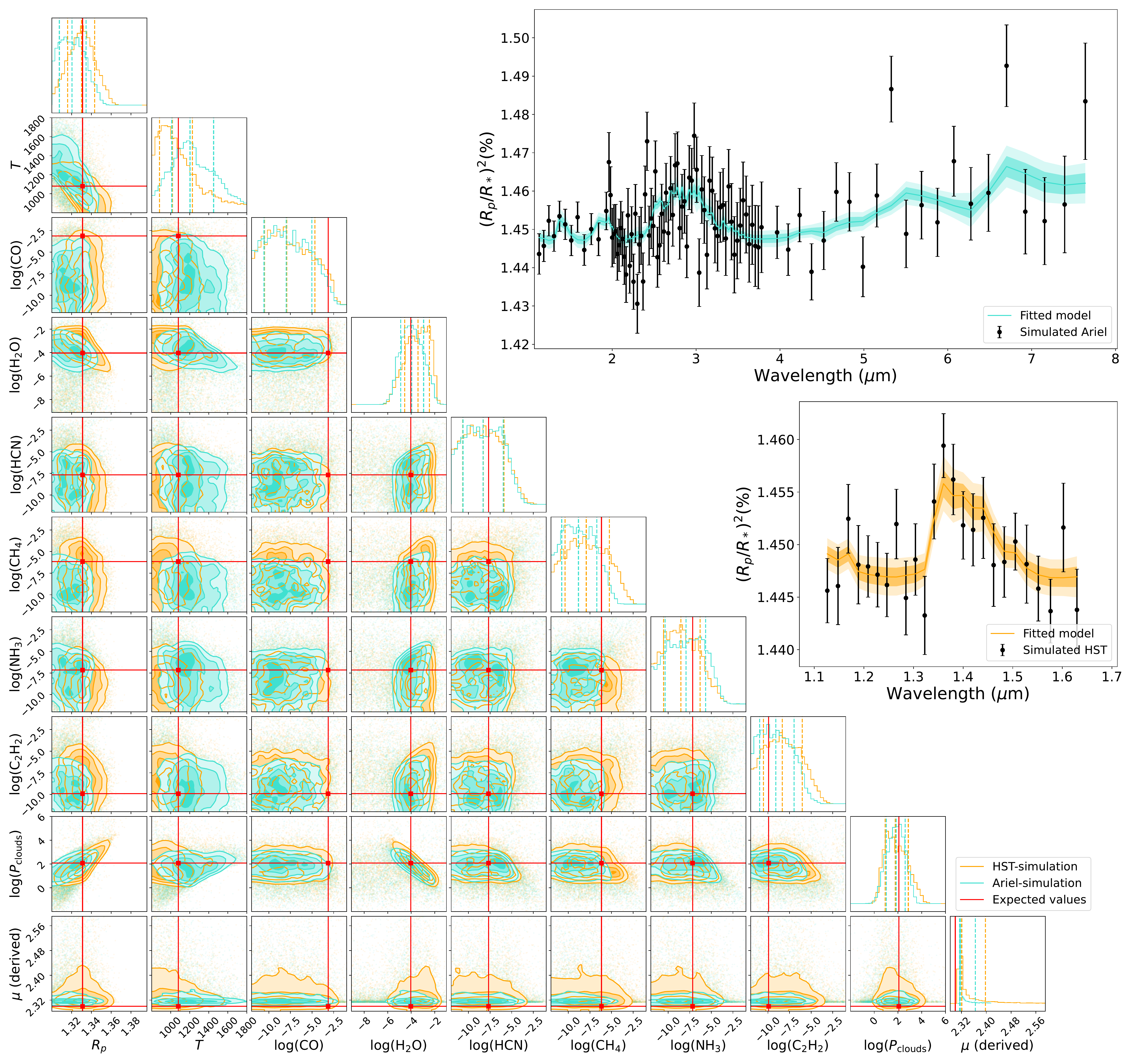}
		\caption{Posterior distributions for the simulated transmission spectrum of HD~209458~b with molecular abundances for a model in thermochemical equilibrium with C/O$\sim$0.9 and log$_{10}$[M/H]=0 (simulation $(a)$) as would be observed by Ariel (in violet), with overplotted the posterior distribution obtained for simulated HST data (in green). The ‘true value' we assumed for each parameter in our TauREx retrieval analysis  is shown in red. Insets: simulated transmission spectra (black points) for Ariel (upper panel) and HST (bottom panel) with overplotted the best fit solution found by TauREx and the correspondent 1$\sigma$ and 2$\sigma$ error bars.}
		\label{fig_sim}
	\end{figure}

\begin{figure}
		\includegraphics[width=\linewidth]{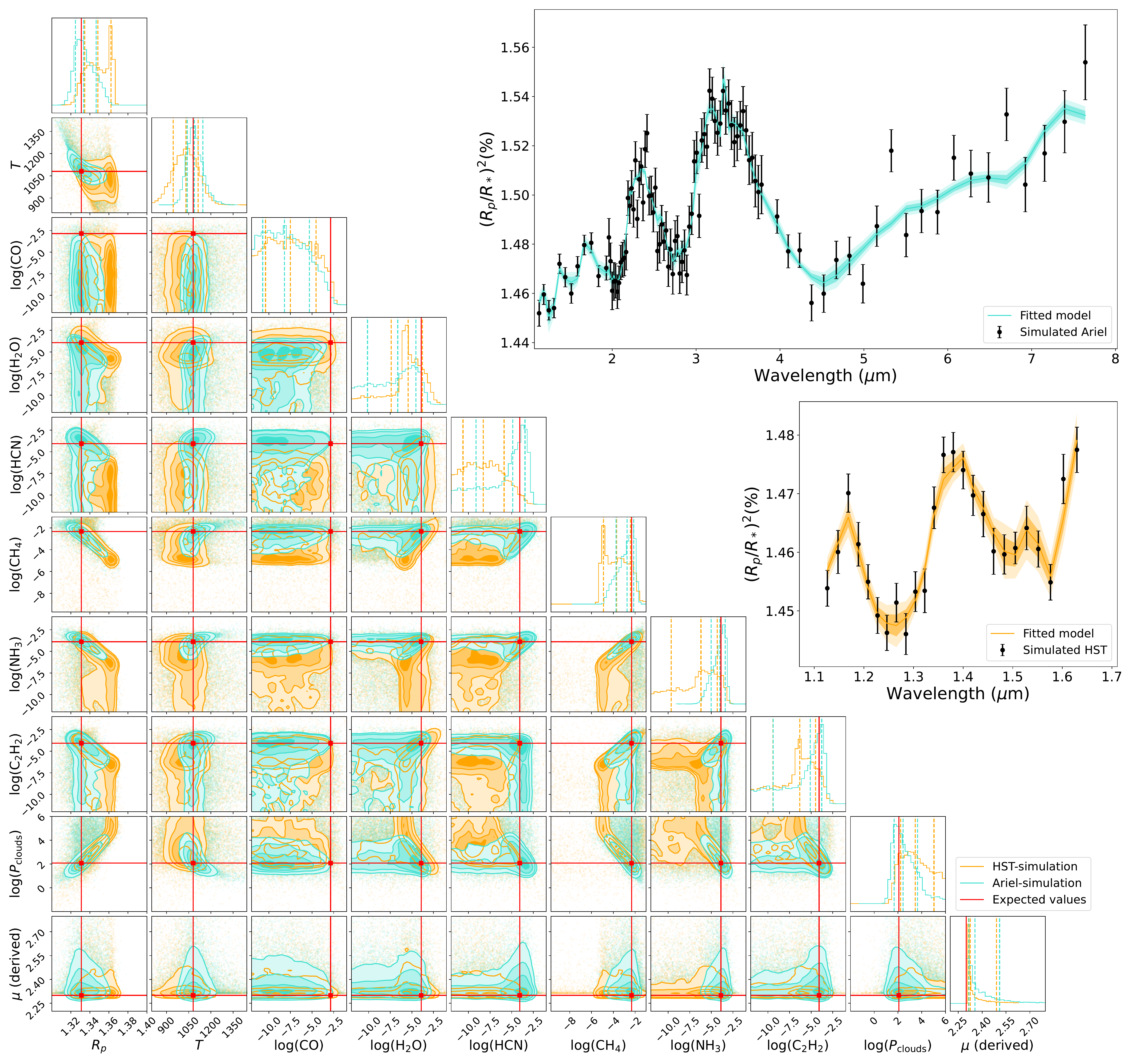}
		\caption{Posterior distributions for the simulated transmission spectrum of HD~209458~b with molecular abundances that maximise the cross-correlation analysis at HRS (simulation $(b)$) as observed by Ariel (in turquoise), with overplotted the posterior distribution obtained for simulated HST data (in orange). The ‘true values' we assumed for each parameter are shown in red. Insets: simulated transmission spectra (black points) for Ariel (upper panel) and HST (bottom panel) with overplotted the best fit solution found by TauREx and the correspondent 1$\sigma$ and 2$\sigma$ error bars.}
		\label{fig_sim_no_clouds}
	\end{figure}

	\noindent{Next}, we binned the high-resolution transmission spectrum to the resolutions of the Ariel spectrometers (i.e. NIRSpec, 1.1–1.95~$\mu$m at a resolving power of R=20, AIRS-Ch0, 1.95–3.9~$\mu$m at R=100, and AIRS-Ch1, 3.9–7.8~$\mu$m at R=30), and we used the instrument noise simulator Ariel Radiometric Model (ArielRad) \citep{Mugnai} to provide a realistic noise model.  Finally, we performed atmospheric retrievals using TauREx3 in fitting mode. 
	Table~\ref{Table5} lists the TauREx retrieval results we obtained, whilst the posterior distribution and the best-fit spectrum are plotted in turquoise in Fig.~\ref{fig_sim} (for the $(a)$ scenario) and in Fig.\ref{fig_sim_no_clouds} (for the $(b)$ simulation). 
	As a comparison, Figs.~\ref{fig_sim}, and~\ref{fig_sim_no_clouds}  show also the posterior distribution and the best-fit transmission spectrum we obtained for simulated HST data. More precisely, we binned our simulated atmospheric spectrum at the WFC3/HST resolution and we perturbed it with the noise obtained in the previous section for the real HST data (right middle box).\\ 
	\begin{itemize}
		\item \textbf{Simulation \textbf{$(a)$}}\\
		Fig.~\ref{fig_sim} shows the retrieval results we obtained for both HST and Ariel observations for the $(a)$ scenario. Only the water vapour posterior distribution is well constrained, whereas we are not able to constrain the CO abundance, and we can only put an upper limit of $\sim$10$^{-3}$ on the amount of the other molecules. The 10~bar radius correlates with the temperature distribution: the
		higher the temperature is, the smaller radius. The Ariel retrieval seems to prefer a slightly higher temperature -and thus a lower 10~bar radius- than the HST simulation. 
		
		\item\textbf{Simulation \textbf{$(b)$}}\\
		The results obtained for the scenario $(b)$ are instead shown in Fig.\ref{fig_sim_no_clouds}. Now that the simulated molecular abundances are higher than before, we can appreciate some differences between Ariel and HST/WFC3. As in the previous simulation, we are not able to constrain the CO abundance distribution. CH$_4$, C$_2$H$_2$, NH$_3$ and HCN  are better constrained in the Ariel simulation.
	\end{itemize}
	These simulations highlight several aspects, which are important to discuss. 
	We can appreciate the improvement in putting constraints on the chemical abundances that we can obtain with Ariel. Indeed, while there is still considerable degeneracy for retrievals from HST/WFC3 spectra, Ariel simulations in Fig.~\ref{fig_sim_no_clouds} show constrained posterior distributions for most of the active gases. This aspect has been also pointed out by \citet{Tinetti2018}. Our simulation highlights how the three Ariel spectrometers represent a step forward compared to the current HST/WFC3. On one side, thanks to its wide spectral coverage, Ariel will permit the detection of several molecular species that do not have strong absorption bands in the WFC3 wavelength range. On the other side, Ariel observations will probe a much wider pressure range than WFC3, from approximately 1 to 10$^3$~mbar \citep{Tinetti2018}.\\ 
	However, when dealing with low molecular abundances, such as those tested in simulation $(a)$, our experiments revealed that also Ariel has some limitations to well constrain the investigated trace-gas abundances and we have to use a combination of HRS results and Ariel data. Indeed, if we consider the Ariel results alone, we could not claim the presence of CO and we can only put an upper limit on the abundance of several tracers. However, combining HRS and Ariel can improve the knowledge of a planet's atmosphere by detecting extra molecules than if the datasets are considered separately.

	\section{Discussion}
	\label{Conclusion}
	\begin{table}
		\label{Table5}
		\centering
		\begin{tabular}{l | c c | c c }	
			\hline\hline
			
			\textbf{Parameters} & \multicolumn{2}{c}{\textbf{simulation (a)}}  & \multicolumn{2}{c}{\textbf{simulation (b)}} \\ 
			\hline 
			& \textbf{Ariel}  & \textbf{HST} & \textbf{Ariel}  & \textbf{HST} \\
			\hline
			log$_{10}$(H$_2$O) & 3.96 $^{+ 1.01 } _{- 0.94 }$ & 3.43   $^{+0.98} _{-1.13}$ &  6.63 $^{+ 2.10 } _{- 3.50 }$   & 5.44 $^{+ 1.65 } _{- 1.95 }$ \\
			log$_{10}$(CO) & 7.99 $^{+ 2.96 } _{- 2.57 }$ & 7.93 $^{+ 3.26 } _{- 2.56 }$ & 8.19 $^{+ 2.87 } _{- 2.52 }$  & 7.52 $^{+ 2.99 } _{- 2.86 }$\\
			log$_{10}$(HCN) & 8.32 $^{+ 2.34 } _{- 2.32 }$& 8.30  $^{+2.41} _{ -2.37 }$ &  4.88 $^{+ 1.42 } _{- 4.25 }$ & 8.29 $^{+ 2.41 } _{- 2.34 }$\\
			log$_{10}$(CH$_4$) & 8.83 $^{+ 2.12 } _{- 1.94 }$& 7.92 $^{+ 2.68 } _{- 2.46 }$ &  2.74 $^{+ 0.61 } _{- 1.01 }$   & 3.70 $^{+ 1.28 } _{- 1.20 }$\\
			log$_{10}$(NH$_3$) & 7.92 $^{+ 2.26 } _{- 2.42 }$ & 8.51 $^{+ 2.21 } _{- 2.02 }$  & 4.00 $^{+ 0.60 } _{- 1.01 }$ & 6.20 $^{+ 1.87 } _{- 3.45 }$ \\
			log$_{10}$(C$_2$H$_2$) & 9.08 $^{+ 2.10 } _{- 1.88 }$ & 8.35 $^{+ 2.32 } _{- 2.17 }$ & 5.11 $^{+ 1.33 } _{- 4.28 }$   & 6.31 $^{+ 1.83 } _{- 3.10 }$ \\
			T$_{\rm P}$ (K) & 1203 $^{+ 248 } _{- 192 }$ & 1017  $^{+211 } _{-135 }$ &  1087$^{+ 58 } _{- 48 }$  & 1031 $^{+ 82 } _{- 86 }$ \\
			R$_{\rm P}$ (R$_{\rm J}$) &  1.320 $^{+ 0.014 } _{- 0.013 }$& 1.330   $^{+0.013} _{-0.014}$ &  1.333 $^{+ 0.013 } _{- 0.009 }$ & 1.3484 $^{+ 0.0138 } _{- 0.0139 }$  \\ 
			log$_{10}$(P$_{\rm clouds}$/1~Pa) & 1.85 $^{+ 0.75 } _{- 0.84 }$& 1.77   $^{+1.10} _{-0.85}$ & 2.42 $^{+ 1.22 } _{- 0.74 }$ &3.46 $^{+ 1.57 } _{- 1.23 }$ \\
			$\mu$ (derived) & 2.316 $^{+ 0.020 } _{- 0.002 }$ & 2.322  $^{+0.054} _{-0.008 }$&  2.35 $^{+ 0.09 } _{- 0.03 }$  & 2.32 $^{+ 0.062 } _{- 0.005 }$ \\
			log (EV) & 792.1   & 217.5 &788.2 & 212.2\\
			\hline
		\end{tabular}
	\end{table}
	In the near future, Ariel will allow for detailed characterisation of thousands of exoplanetary atmospheres. In this perspective, exploiting the synergies between Ariel and current (e.g. GIANO-B, SPIROU, CARMENES) and upcoming ground-based high-resolution spectrographs (NIRPS, CRIRES+, and ELTs) represents the future of exoplanetary science. The combination of HRS and LRS can break the degeneracies that arise when we apply the two techniques individually. On one hand, the simultaneous spectra coverage of Ariel will permit us to determine important atmospheric properties as the molecular abundances, the clouds coverage, and the T/P profile. Moreover, it could provide a local pseudo-continuum to HRS observations. On the other hand, the contributions of multiple species are expected to overlap at the low resolution of Ariel, and high-resolution instruments like GIANO-B can help in understanding which atmospheric constituents are expected. Furthermore, in presence of a deep cloud deck, some spectral features that could be masked with LRS, may be accessible with HRS, since it probes higher atmospheric levels. 
	It is thus clear that only a combination of low- and high-resolution data in a consistent retrieval framework
	can provide absolute molecular abundances, avoiding
	confusion between species. The power of having a joint retrieval of multi-spectral resolution data was shown by \citet{Brogi2019} who provided a simple framework for combining HRS data and LRS data within a unified likelihood function. However, these calculation were restricted to an extremely narrow wavelength
	range at high-resolution (2270-2350~nm), and limited to emission spectroscopy data. No such framework has ever been applied to transmission spectroscopy.
	An efficient LRS+HRS retrieval framework could allow to achieve unprecedented atmospheric characterization and thus to 
	(i) better understand what are the main parameters that govern the chemistry and physics of exoplanet's
	atmospheres, and
	(ii) derive the C/O and O/H ratios which provide insights into formation and evolution mechanisms \citep[e.g.][]{Madhusudhan_2014}.\\
	\vspace{0.5cm}\\
	
	\noindent{In this work, we  took HD~209458~b as a reference, but the same study can be performed on several exoplanets. To understand the number of potential objects on which we can perform a similar analysis, it would be useful to know how many of the future Ariel targets could be observed with ground-based HRS. We downloaded the list of transiting planets from the \textit{TEPCat  Catalog} \citep{Southworth_2011}, and for each of them we estimated the atmospheric signal to noise ratio in transmission ($S/N$) under the following assumptions: (i) giant planets with helium-hydrogen dominated atmospheres (R$_{\rm{P}}\geq$ 3.5 R$_{\oplus}$), (ii) no clouds, (iii) and pure photon noise. We calculated the expected $S/N$ as follows:}
	\begin{equation}
		S/N =\frac{2HR_P}{R\star^2}*\sqrt{T}*\sqrt{F},
	\end{equation}
	\noindent{	where we account for the transit duration ($T$), the stellar magnitude in K band ($F=10^{-\frac{magK}{2.5}}$), the atmospheric scale height ($H$), the planetary ($R_P$) and the stellar radius ($R_\star$). We divided each of the estimated S/N to the expected S/N of HD~209458~b.\\
		In Section~\ref{GIANO-B} we reported a detection of water vapour with a statistical confidence of 10~$\sigma$ by considering 4 transits, therefore around 5~$\sigma$ ($\sigma_{HD209}=10~\sigma/\sqrt{4.}$) for each transit. If we assume a statistical significance threshold of 4~$\sigma$ for water detection and to gather 5 transit observations for each target, $\sim$25 targets in our sample of transiting planets could be studied with both Ariel and HRS instruments such as GIANO-B at 4-m class telescope. The number of suitable targets is limited by the 4-m aperture of the TNG, which restricts the observable sample to relatively bright stars. Indeed, most of the planet-hosting stars are remarkably fainter than HD209458b, and thus we are not able to observe them with a sufficiently high S/N.} In the near future, as soon as high-resolution instrumentation at ELTs is available, this sample is expected to increase.     \\
	\\
	
	Even though the analysis presented in this work is focused on the nIR band, we have to highlight also the benefits that can be obtained by combining HRS and Ariel data in the VIS. Studies of hot/ultra-hot Jupiters with the Hubble STIS/WFC3 \citep[e.g.][]{Evan2018} have recently shown indications of excess absorption possibly due to TiO, VO, and FeH or H-opacity. Ariel could measure these excesses but is not able to distinguish between the different opacities due to the spectral bins' width. Thus, HRS could be employed to place upper limits on the molecular abundances. The HRS  prior information could then be incorporated into the retrieval of the Ariel data -in a similar way to that proposed in this study- to help break the degeneracy that will be seen. This approach has recently been taken for the VO molecule. Indeed, HST observations of WASP-121~b seemed to show evidence of VO \citep{Evan2018}, whilst high-resolution data only succeeded in putting an upper limit on this opacity \citep{Merritt2020}. Thus a future combination between optical high-resolution spectrographs (such as HARPS-N@TNG) and Ariel's VIS photometric channels could put constraints also on the TiO/VO/Na opacities.
	\section{Conclusions}
	In this paper, we compared results between HRS and LRS gathered in transmission for HD~209458~b. We applied different methods to analyse the two different datasets. On the one hand, we disentangle the planetary spectrum from the stellar and telluric contamination and the cross-correlation technique to extract the planet's signal from our GIANO-B data. On the other one, we used the Python package \texttt{Iraclis} to extract the transmission spectrum of HD~209458~b from the HST/WFC3 raw images and the algorithm TauRex3 to perform retrieval analysis. 
	We noted a considerable degeneracy in the posteriors distributions of many molecular species obtained at low resolution.\\
	Successively, we performed a simulation to test the ability of the Ariel space mission to give precise constraints on the atmospheric chemical abundances. 
	By using TauRex3 in the forward mode, we simulated a transmission spectrum for HD~209458~b containing H$_2$O, CO, CH$_4$, NH$_3$, C$_2$H$_2$,and HCN. We added instrumental noise to the model by using the ArielRad simulator and we interpreted it with TauREx3 in retrieval mode. Our simulation showed how, unlike HST/WFC3, the retrieved atmospheric information we could obtain from Ariel is much better constrained, and how much the exoplanetary community could benefit from exploiting the synergy between Ariel, and current or upcoming HRS ground-based telescopes.
	The work presented in this paper set only the basis to combine LRS and HRS. In this respect, we need to further stress that in this work we did not carry out a true combination of LRS and HRS results such as that advocated in recent literature \citep[e.g.,][] {Brogi2017,Brogi2019} and based on a statistically supported method to combine datasets, but rather we selected the chemical absorbers to be investigated at low-resolutions based on HRS results, and for each molecule we used as starting value of the TauREx retrieval the abundance that  maximises the detection significance at HRS. However, to perform a true combination of these two techniques, work still needs to be done. The natural starting point will be the effective combination of HST/WFC3 and GIANO-B spectra by maximizing a total likelihood which is the combination of a low-resolution likelihood
	\citep{Madhu2009} and an high-resolution likelihood \citep{Brogi2019}, over much broader wavelength coverage than that used in the pioneering work of \cite{Brogi2019}. Such a framework will i) yield unprecedented constraints on both the atmospheric composition of the studied exoplanets and their formation/migration mechanisms and, even more importantly, ii) constitute a reference for combined analyses of future low-resolution data gathered with  JWST and/or ARIEL and ground-based high-resolution spectra, including those that will be acquired with future ELTs.

	\begin{acknowledgements}
		We thank Lorenzo Mugnai for the help he gave us with his ArielRad code. 
		We want to thank the anonymous referees for the constructive
		comments that helped improve the quality of the manuscript. GG acknowledges the financial support of the 2017 PhD fellowship programme of INAF. PG gratefully acknowledge support from the Italian Space Agency (ASI) under contract 2018-24-HH.0. We acknowledge financial contribution from the agreement ASI-INAF n.2018-16-HH.0.
		We acknowledge the support of the Ariel ASI-INAF agreement n. 2018-22-HH.0. Data presented in this paper were obtained from the Mikulski Archive for Space Telescopes (MAST). We acknowledge the PI of the 12181 program Drake Deming.
	\end{acknowledgements}

	\bibliographystyle{spbasic}
	\bibliography{bib_.bib}
\end{document}